# Influence of a momentum dependent interaction on the isospin dependence of fragmentation and dissipation in intermediate energy heavy ion collisions

Jian-Ye Liu,[1,2,3] Wen-Jun Guo,[2] Yong Zhong Xing,[2,4] Wei Zou,[1,2] and Xi-Guo Lee[1,2]

[1]*Center of Theoretical Nuclear Physics, National Laboratory of Heavy Ion Accelerator, Lanzhou 730000, People's Republic of China*
[2]*Institute of Modern Physics, Chinese Academy of Sciences, P.O. Box 31, Lanzhou 730000, People's Republic of China*
[3]*Department of Chemistry and Physics, P.O. Box 419, Arkansas State University, State University, Arkansas 72467-0419*
[4]*Department of Physics, Tianshui Normal College, Gansu Tianshui 741000, People's Republic of China*

We studied the influence of a momentum dependent interaction in the context of isospin effects on fragmentation and dissipation in intermediate energy heavy ion collisions by using an isospin dependent quantum molecular dynamics model. It is shown that nuclear stopping, the number of nucleons emitted, and the multiplicity of intermediate mass fragments are larger with a momentum dependent interaction than without. In particular, the differences for these observables, when using an isospin dependent in-medium nucleon-nucleon cross section versus an isospin independent one, are also larger at high energies for a momentum dependent interaction than without one. Therefore, momentum dependence enhances the sensitivities of those observables to the isospin effect of the in-medium nucleon-nucleon cross section towards high beam energies.

## I. INTRODUCTION

The rapid progress in producing energetic radioactive beams has offered an excellent opportunity to investigate various isospin effects in the dynamics of nuclear reactions [1–5]. This study has made it possible to obtain crucial information about the equation of state of isospin asymmetric nuclear matter and about the isospin dependent in-medium nucleon-nucleon cross section. This information is important for understanding both novel properties of neutron- or proton-rich nuclei as well as explosion mechanisms of supernovae and the cooling rates of protoneutron stars. Recently, several interesting isospin effects in heavy ion collisions were explored both experimentally [6–17] and theoretically [18–32]. For a recent review, see Ref. [1]. However, two essential ingredients in heavy ion collision dynamics, the symmetry potential of the mean field and the isospin dependent in-medium nucleon-nucleon cross section, are still not well determined. Moreover, the nonlocal property of the nuclear interaction leads to a repulsive momentum dependent interaction in the dynamics of intermediate energy heavy ion collisions. Even though the roles of momentum dependent interactions in heavy ion collisions have been studied for many years [33–38], the underlying influence of momentum dependence on isospin effects in intermediate energy heavy ion collisions is poorly known.

Based on an isospin dependent quantum molecular dynamics model we study in this work the underlying influence of a momentum dependent interaction (MDI) on isospin effects in intermediate energy heavy ion collisions. We find pronounced isospin effects in the nuclear stopping $R$, the multiplicity of intermediate mass fragments $N_{imf}$, and the number of nucleons emitted $N_n$ ($N_p$) due mainly to the isospin dependent in-medium nucleon-nucleon cross section in the high energy region [39–41]. Furthermore, we investigate the influence of momentum dependence on isospin effects in fragmentation and dissipation with (MDI) and without (no-MDI) a momentum dependent interaction in the high energy region. It is found that a momentum dependent interaction enhances the sensitivities of nuclear stopping, the multiplicity of intermediate mass fragments, and the number of nucleons emitted to the isospin effect on in-medium nucleon-nucleon cross sections.

## II. THEORETICAL MODEL

A quantum molecular dynamics (QMD) model contains two dynamical ingredients, the density dependent mean field and the in-medium nucleon-nucleon cross section. In order to describe the isospin dependence appropriately, the QMD model must be modified properly. The density dependent mean field must contain correct isospin terms, including the symmetry potential and the Coulomb potential. The in-medium nucleon-nucleon cross section should be different for neutron-neutron (or proton-proton) and neutron-proton collisions, in which Pauli blocking should be included by distinguishing between neutrons and protons. In addition, the initial condition of the ground states of two colliding nuclei must also contain isospin effects.

Considering the above ingredients, we have made important modifications to QMD to obtain an isospin dependent quantum molecular dynamics (IQMD) approach [3,35]. The IQMD code consists of two parts, an initialization step and the calculation of the reaction dynamics itself. For the initialization, the code IQMD is used to determine the ground state properties of the colliding nuclei, such as binding energies and root-mean-square (rms) radii, so that they agree with experimental data. The initial density distributions of the colliding nuclei in IQMD are obtained from the Skyrme-Hartree-Fock model with parameter set SKM* [42]. The interaction potential is

$$U(\rho) = U^{Sky} + U^c + U^{sym} + U^{Yuk} + U^{MDI} + U^{Pauli}, \quad (1)$$

where $U^c$ is the Coulomb potential. The density dependent Skyrme potential $U^{Sky}$, the Yukawa potential $U^{Yuk}$, the mo-mentum dependent interaction $U^{\text{MDI}}$, and the Pauli potential $U^{\text{Pauli}}$ [43,44] are given by the following equations:

$$U^{\text{Sky}} = \alpha\left(\frac{\rho}{\rho_0}\right) + \beta\left(\frac{\rho}{\rho_0}\right)^\gamma, \quad (2)$$

$$U^{\text{Yuk}} = t_3 \exp\left(\frac{|\vec{r}_1-\vec{r}_2|}{m}\right) \Big/ \frac{|\vec{r}_1-\vec{r}_2|}{m}, \quad (3)$$

$$U^{\text{MDI}} = t_4 \ln^2[t_5(\vec{p}_1-\vec{p}_2)^2+1]\frac{\rho}{\rho_0}, \quad (4)$$

and

$$U^{\text{Pauli}} = V_p\left\{\left(\frac{\hbar}{p_0 q_0}\right)^3 \exp\left(-\frac{(\vec{r}_i-\vec{r}_j)^2}{2q_0^2}-\frac{(\vec{p}_i-\vec{p}_j)^2}{2p_0^2}\right)\right\}\delta_{p_i p_j}, \quad (5)$$

with

$$\delta_{p_i p_j} = \begin{cases} 1 & \text{for neutron-neutron or proton-proton} \\ 0 & \text{for neutron-proton.} \end{cases}$$

TABLE I. The parameters of the interaction potential.

| | $\alpha$ (MeV) | $\beta$ (MeV) | $\gamma$ | $t_3$ (MeV) | $m$ (fm) | $t_4$ (MeV) | $t_5$ (MeV$^{-2}$) | $V_p$ (MeV) | $p_0$ (MeV/$c$) | $q_0$ (fm) |
|---|---|---|---|---|---|---|---|---|---|---|
| MDI | −390.1 | 320.3 | 1.14 | 7.5 | 0.8 | 1.57 | $5\times 10^{-4}$ | 30 | 400 | 5.64 |
| no-MDI | −356 | 303 | 1.1667 | 7.5 | 0.8 | 0.0 | 0.0 | 30 | 400 | 5.64 |

We used the following four different forms of the symmetry potential [2,3]:

$$U_1^{\text{sym}} = cu\,\delta\tau_z, \quad (6)$$

$$U_2^{\text{sym}} = cu^2\delta\tau_z + \frac{1}{2}cu^2\delta^2, \quad (7)$$

$$U_3^{\text{sym}} = \pm cu^{1/2}\delta - \frac{1}{4}cu^{1/2}\delta^2, \quad (8)$$

$$U_0^{\text{sym}} = 0, \quad (9)$$

where

$$\tau_z = \begin{cases} 1 & \text{for neutron} \\ -1 & \text{for proton,} \end{cases}$$

and $c$ is the strength of the symmetry potential chosen to be 32 MeV. In this work, $u=\rho/\rho_0$ is the reduced density and $\delta=\rho_n-\rho_p/\rho_n+\rho_p=\rho_n-\rho_p/\rho$ is the relative neutron excess. $\rho$, $\rho_0$, $\rho_n$, and $\rho_p$ are the total, normal, neutron, and proton densities, respectively. To investigate the influence of momentum dependence on the fragmentation and dissipation we calculated both observables, with (MDI) and without (no-MDI) a momentum dependent interaction. The parameters of the interaction potentials are given in Table I.

It is worth mentioning that recent studies of collective flow in heavy ion collisions at intermediate energy have indicated a reduction of in-medium nucleon-nucleon cross sections. An empirical, in-medium density dependent nucleon-nucleon cross section [27] has been suggested as follows:

$$\sigma_{NN}^{\text{med}} = \left(1+\alpha\frac{\rho}{\rho_0}\right)\sigma_{NN}^{\text{free}}, \quad (10)$$

where $\sigma_{NN}^{\text{free}}$ is the experimental nucleon-nucleon cross section [45]. The above expression with $\alpha\approx -0.2$ has been found to reproduce well the flow data. The free neutron-proton cross section is about a factor of 3 larger than the free proton-proton or the free neutron-neutron cross section below 400 MeV, which contributes the main isospin effect from nucleon-nucleon collisions in heavy ion collisions. In fact, the ratio of neutron-proton cross section to proton-proton (or neutron-neutron) cross section in the medium, $\sigma_{np}/\sigma_{pp}$, depends sensitively on the evolution of the nuclear density distribution and the beam energy. We used Eq. (10) to take into account medium effects, in which the neutron-proton cross section is always larger than the neutron-neutron or proton-proton cross section at the beam energies considered in this paper. We constructed the clusters by means of a modified coalescence model [35], in which coalescence occurs when the relative particle momentum is smaller than $p_0 = 300$ MeV/$c$ and the relative distance is smaller than $R_0 = 3.5$ fm. The restructured aggregation model [46] has been applied to avoid nonphysical clusters. The main physics ingredients and their numerical realization in the IQMD model can be found in Refs. [3,29,33–35].

### III. RESULTS AND DISCUSSIONS

The isospin effect of the in-medium nucleon-nucleon cross section on the observables is defined by the difference between the observable for an isospin dependent nucleon-nucleon cross section $\sigma^{\text{iso}}$ and that for an isospin independent nucleon-nucleon cross section $\sigma^{\text{no-iso}}$ in the medium. Here $\sigma^{\text{iso}}$ means $\sigma_{np} \geqslant \sigma_{nn} = \sigma_{pp}$ and $\sigma^{\text{no-iso}}$ means $\sigma_{np} = \sigma_{nn} = \sigma_{pp}$, where $\sigma_{np}$, $\sigma_{nn}$, and $\sigma_{pp}$ are the neutron-proton, neutron-neutron, and proton-proton in-medium cross sections, respectively.

To study the influence played by the MDI on isospin effects of the fragmentation and dissipation in intermediate energy heavy ion collisions we investigated the time evolutions of the following variations on the nuclear stopping $R$, the multiplicity of intermediate mass fragment $N_{\text{imf}}$, and the number of nucleon emissions $N_n(N_p)$ due to isospin effects from the in-medium nucleon-nucleon cross section with (MDI) and without (no-MDI) a momentum dependent interaction. They are indicated by $\Delta R(\Delta\sigma,U_2^{\text{sym}})$, $\Delta N_{\text{imf}}(\Delta\sigma,U_1^{\text{sym}})$, $\Delta N_n(\Delta\sigma,U_1^{\text{sym}})$, and $\Delta N_p(\Delta\sigma,U_1^{\text{sym}})$, respectively. It is more quantitative to define

$$\Delta R^{\text{MDI(no-MDI)}}(\Delta\sigma,U_2^{\text{sym}})$$
$$\equiv R^{\text{MDI(no-MDI)}}(\sigma^{\text{iso}},U_2^{\text{sym}})$$
$$- R^{\text{MDI(no-MDI)}}(\sigma^{\text{no-iso}},U_2^{\text{sym}}), \quad (11)$$

$$\Delta N_{\text{imf}}^{\text{MDI(no-MDI)}}(\Delta\sigma,U_1^{\text{sym}})$$
$$\equiv N_{\text{imf}}^{\text{MDI(no-MDI)}}(\sigma^{\text{iso}},U_1^{\text{sym}})$$
$$- N_{\text{imf}}^{\text{MDI(no-MDI)}}(\sigma^{\text{no-iso}},U_1^{\text{sym}}), \quad (12)$$

$$\Delta N_n^{\text{MDI(no-MDI)}}(\Delta\sigma,U_1^{\text{sym}})$$
$$\equiv N_n^{\text{MDI(no-MDI)}}(\sigma^{\text{iso}},U_1^{\text{sym}})$$
$$- N_n^{\text{MDI(no-MDI)}}(\sigma^{\text{no-iso}},U_1^{\text{sym}}), \quad (13)$$

$$\Delta N_p^{\text{MDI(no-MDI)}}(\Delta\sigma,U_1^{\text{sym}})$$
$$\equiv N_p^{\text{MDI(no-MDI)}}(\sigma^{\text{iso}},U_1^{\text{sym}})$$
$$- N_p^{\text{MDI(no-MDI)}}(\sigma^{\text{no-iso}},U_1^{\text{sym}}). \quad (14)$$

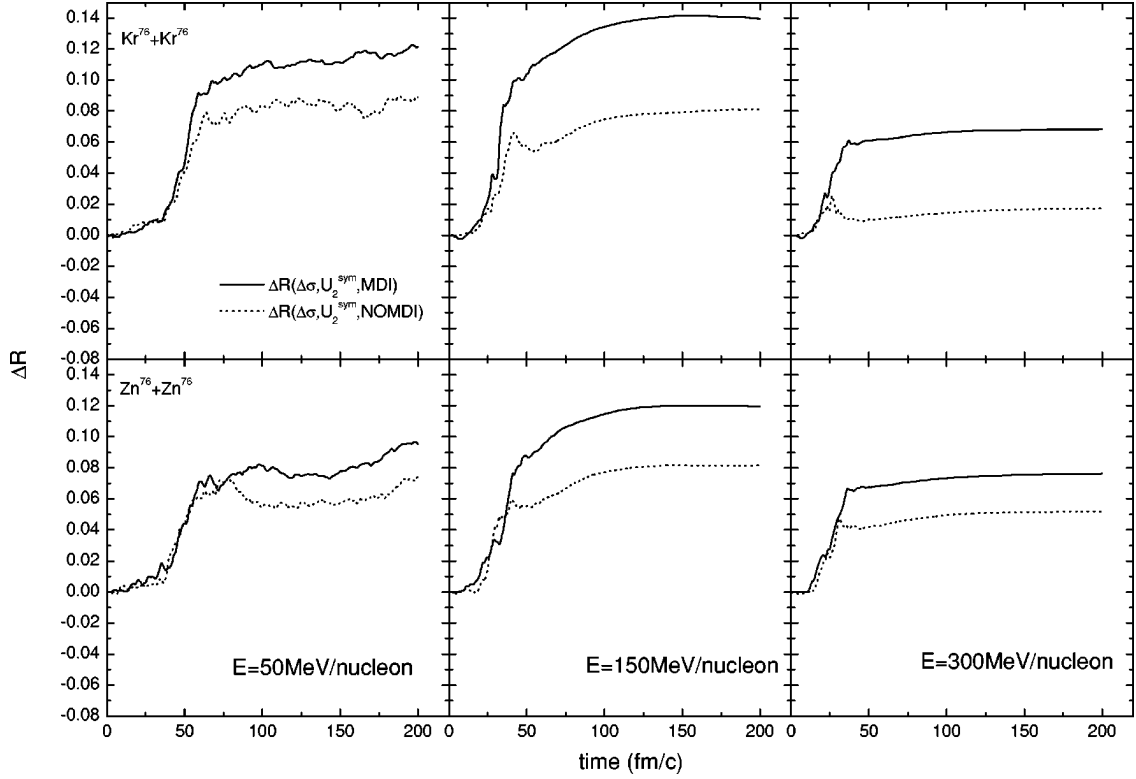

FIG. 1. Time evolution of the variation of nuclear stopping $\Delta R$ due to the isospin effect from the in-medium nucleon-nucleon cross section for head-on collisions of $^{76}$Kr+$^{76}$Kr (top panel) and $^{76}$Zn+$^{76}$Zn (bottom panel) at beam energies of 50, 150, and 300 MeV/nucleon with (MDI, solid lines) and without (no-MDI, dotted lines) a momentum dependent interaction.

$U_2^{\text{sym}}$ in Eq. (11) and $U_1^{\text{sym}}$ in Eqs. (12), (13), and (14) are the symmetry potentials used for calculating the corresponding variations.

To describe the nuclear stopping in heavy ion collisions we used the transverse-parallel ratio of the momentum defined by $R=(2/\pi)(\Sigma_i^A|P_\perp(i)|)/(\Sigma_i^A|P_\parallel(i)|)$, where the total mass $A$ is the sum of the projectile mass $A_p$ and the target mass $A_t$. The transverse and the parallel components of the momentum of the $i$th nucleon are $P_\perp(i)=\sqrt{P_x(i)^2+P_y(i)^2}$ and $P_{//}(i)=P_z(i)$, respectively. The multiplicity of the intermediate mass fragments is defined as the number of fragments with charge numbers from 4 to 18.

Figure 1 shows the time evolution of the variations $\Delta R^{\text{MDI}}(\Delta\sigma,U_2^{\text{sym}})$ (solid lines) and $\Delta R^{\text{no-MDI}}(\Delta\sigma,U_2^{\text{sym}})$ (dotted lines) for systems $^{76}$Kr+$^{76}$Kr (top panels) and $^{76}$Zn +$^{76}$Zn (bottom panels) at $E=50$ MeV/nucleon (left panels), 150 MeV/nucleon (middle panels), and 300 MeV/nucleon (right panels) and $b=0.0$ fm. From Fig. 1 it is clear that all of the solid lines are always higher than the dotted lines, i.e., the MDI enhances the sensitivity of the nuclear stopping $R$ to the isospin effect of the in-medium nucleon-nucleon cross section.

We also show the time evolution of the variations $\Delta N_{\text{imf}}$ due to the isospin dependence of the in-medium nucleon-nucleon cross section with (MDI) and without (no-MDI) momentum dependent interaction in Fig. 2 for the systems $^{58}$Fe+$^{58}$Fe (left panel) at $E=100$ MeV/nucleon and $b =4.0$ fm as well as for $^{112}$Sn+$^{40}$Ca (right panel) at $E$ $=80$ MeV/nucleon and $b=0.0$ fm. All $\Delta N_{\mathrm{imf}}^{\mathrm{MDI}}(\Delta\sigma,U_1^{\mathrm{sym}})$ (solid lines) are higher than the corresponding $\Delta N_{\mathrm{imf}}^{\mathrm{no\text{-}MDI}}(\Delta\sigma,U_1^{\mathrm{sym}})$ (dotted lines) for the two systems. This means that the momentum dependence also enhances the sensitivity of $N_{\mathrm{imf}}$ to the isospin effect from the in-medium nucleon-nucleon cross section.

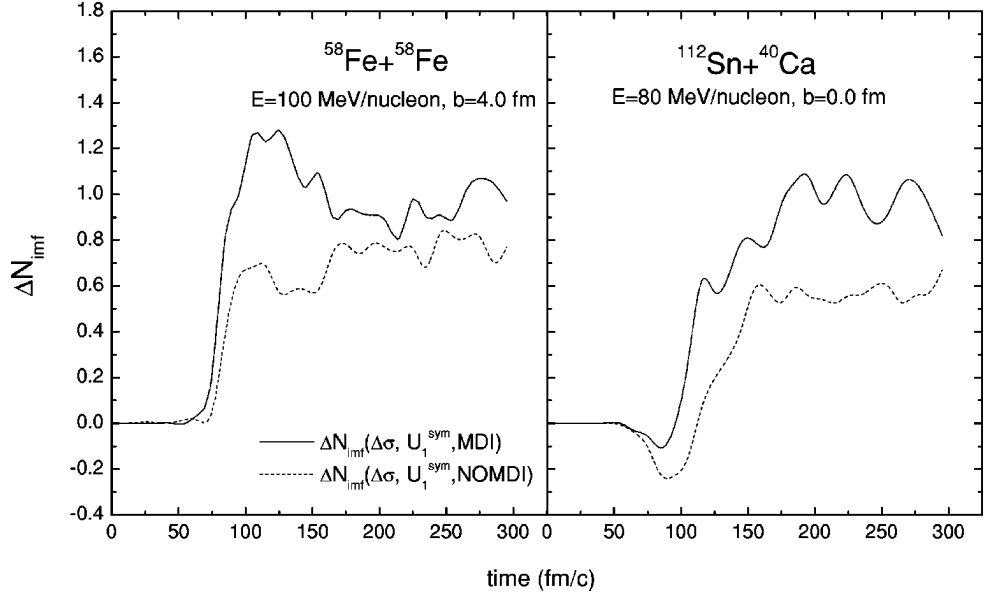

FIG. 2. The time evolution of the variation of $\Delta N_{\mathrm{imf}}$ due to the isospin effect from the in-medium nucleon-nucleon cross section with (MDI, solid lines) and without (no-MDI, dotted lines) a momentum dependent interaction for the colliding systems $^{58}$Fe+$^{58}$Fe (left panel) at $E=100$ MeV/nucleon, $b=4.0$ fm, and $^{112}$Sn+$^{40}$Ca (right panel) at $E=80$ MeV/nucleon, $b=0.0$ fm.

In Fig. 3 we show the time evolution of the variations $\Delta N_n$ of the number of emitted neutrons due to the isospin effect from the in-medium nucleon-nucleon cross section with (MDI, solid lines) and without (no-MDI, dotted lines) a momentum dependent interaction. The left part of Fig. 3 shows the results for the system $^{76}$Kr+$^{40}$Ca at $E=150$ MeV/nucleon and $b=4$ fm. The right part shows the corresponding results for proton emission. From Fig. 3 one can see that MDI enhances the sensitivities of $N_p$ and $N_n$ to the isospin effect from the in-medium nucleon-nucleon cross section similar to $N_{\mathrm{imf}}$ and $R$.

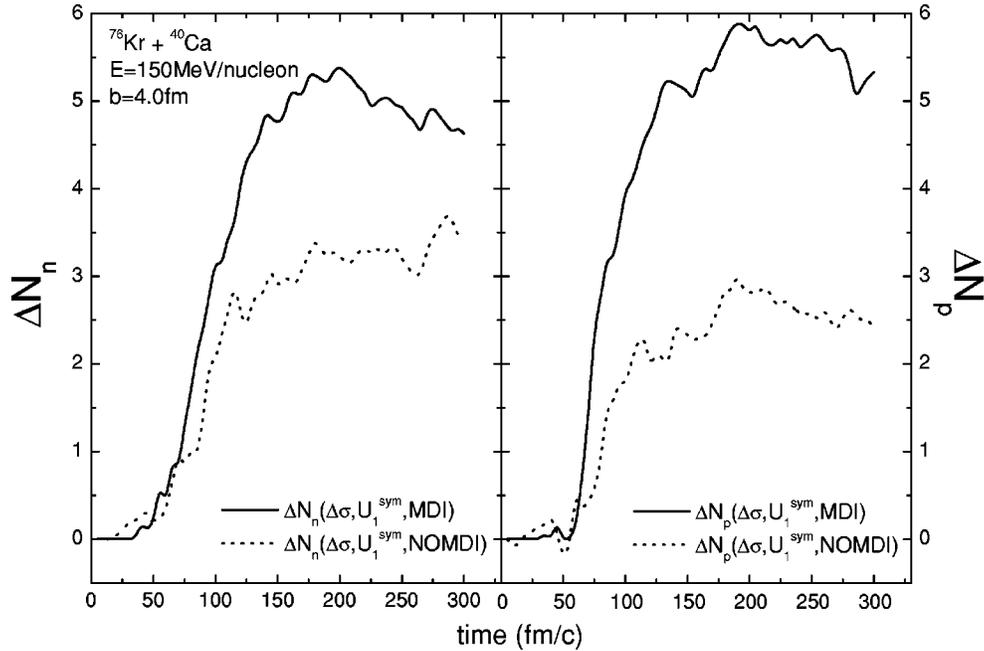

FIG. 3. The time evolution of the variation of the number of emitted neutrons $\Delta N_n^{\mathrm{MDI(no\text{-}MDI)}}(\Delta\sigma,U_1^{\mathrm{sym}})$ due to the isospin effect from the in-medium nucleon-nucleon cross section with (MDI, solid line) and without (no-MDI, dotted line) a momentum dependent interaction (left panel). The right panel is the same as the left panel but for emitted protons at $E=150$ MeV/nucleon and $b=4$ fm for the system $^{76}$Kr+$^{40}$Ca.

To summarize, it is clear that a momentum dependent interaction enhances the sensitivities of $R$, $N_{\mathrm{imf}}$, $N_p$, and $N_n$ to the isospin effect from the in-medium $NN$ cross section. A momentum dependent interaction deflects the nucleons in the transverse direction during the initial phase of the reaction when two nuclei with large relative momentum penetrate each other. This produces a larger momentum damping and thus leads to more dissipation and fragmentation during the reactions, than when the interaction does not depend on the momentum. Therefore, with a momentum dependent interaction the observables $R$, $N_{\mathrm{imf}}$, and $N_p(N_n)$ are more sensitive to the isospin effect of the in-medium nucleon-nucleon cross section in high energy heavy ion collisions.

## IV. SUMMARY AND CONCLUSIONS

We have studied the influence of a momentum dependent interaction on the isospin effects of fragmentation and dissipation from mainly the in-medium nucleon-nucleon cross section by using the IQMD model. The results show that the nuclear stopping, the multiplicity of intermediate mass fragments, and the number of emitted nucleons with the momentum dependent interaction are always larger than those without the momentum dependent interaction in the energy region studied here. In particular, the variations due to the isospin effect from the in-medium nucleon-nucleon cross section with the momentum dependent interaction are greater than those without the momentum dependent interaction. Therefore, the momentum dependent interaction enhances the sensitivities of these observables to the isospin effect from the in-medium nucleon-nucleon cross sections in heavy ion collisions towards high beam energies.

## ACKNOWLEDGMENTS

We thank Professor Bao-An Li for helpful discussions. This work was supported by the Major State Basic Research Development Program in China Under Contract No. G2000077400, ''One Hundred Person Project'' of the Chinese Academy of Sciences, the National Natural Sciences Foundation of China under Grant Nos. 10175080, 10004012, 10175082, and The CAS Knowledge Innovation Project No. KJCX2-SW-N02.